\documentstyle[12pt]{article}
\newcommand{\be}{\begin{equation}}
\newcommand{\ee}{\end{equation}}

\newcommand{\bear}{\begin{eqnarray}}
\newcommand{\eear}{\end{eqnarray}}

\def\fun#1#2{\lower3.6pt\vbox{\baselineskip0pt\lineskip.9pt
        \ialign{$\mathsurround=0pt#1\hfill##\hfil$\crcr#2\crcr\sim\crcr}}}

\renewcommand\({\left(}
\renewcommand\){\right)}
\renewcommand\[{\left[}

\newcommand\eq[1]{Eq.~(\ref{#1})}

\newcommand\eea{\end{eqnarray}}
\newcommand\bea{\begin{eqnarray}}
        		
\newcommand\GeV{\,\mbox{GeV}}
\newcommand\MeV{\,\mbox{MeV}}

\newcommand\mpl{M_{\rm Pl}}

\newcommand\lsim{\mathrel{\rlap{\lower4pt\hbox{\hskip1pt$\sim$}}
    \raise1pt\hbox{$<$}}}
\newcommand\gsim{\mathrel{\rlap{\lower4pt\hbox{\hskip1pt$\sim$}}
    \raise1pt\hbox{$>$}}}

\def\dslash{\not{\hbox{\kern-2pt $\partial$}}}
\def\Dslash{\not{\hbox{\kern-4pt $D$}}}
\def\Oslash{\not{\hbox{\kern-4pt $O$}}}
\def\Qslash{\not{\hbox{\kern-4pt $Q$}}}
\def\pslash{\not{\hbox{\kern-2.3pt $p$}}}
\def\kslash{\not{\hbox{\kern-2.3pt $k$}}}
\def\qslash{\not{\hbox{\kern-2.3pt $q$}}}

 \newtoks\slashfraction
 \slashfraction={.13}
 \def\slash#1{\setbox0\hbox{$ #1 $}
 \setbox0\hbox to \the\slashfraction\wd0{\hss \box0}/\box0 }
 
\newcommand\sub[1]{_{\rm #1}}

\newcommand\vpq{F\sub{PQ}}

\begin{document}

\begin{titlepage}

\title{ {\bf Flatons  and Peccei-Quinn Symmetry}
\vskip-4.cm
\rightline{ {\normalsize IEM--FT--12x/99}   }  \vskip-1ex
\rightline{ {\normalsize INFNFE-02-99}    }  \vskip-1ex
\rightline{ {\normalsize KIAS--P99016}  }  
\vskip2cm
}

\author{
{\bf E. J. Chun }\\   
Korea Institute for Advanced Study\\
207-43 Cheongryangri-dong, Dongdaemun-gu, Seoul 130-012, Korea  \\[1ex]
{\bf D. Comelli}\\
INFN, sezione di Ferrara\\
Ferrara, via Paradiso 12 44100 Italy\\[1ex]
{\bf David H. Lyth }\\
Department of Physics,\\
Lancaster University,\\
Lancaster LA1 4YB.~~~U.~K.}

\date{March 1999} 
\maketitle
\def\baselinestretch{1.15}
\begin{abstract}
We study in detail a supersymmetric
Peccei-Quinn model, which has a DFSZ and a  KSVZ version.
The fields breaking the Peccei-Quinn (PQ) symmetry 
correspond to flat directions (flaton fields) and have unsuppressed couplings
when PQ symmetry is unbroken.
The models have interesting particle physics phenomenology. The
PQ scale is 
naturally generated through radiative corrections; also, in the 
 DFSZ case, the 
 $\mu$ problem can be solved and 
 neutrino masses can be generated.
Cosmologically they  lead to  a short period of 
thermal inflation making the axion an 
excellent dark matter candidate if one of the flaton  fields
has a positive effective mass-squared at early times
 but with too low a 
density in the opposite case.
 A highly relativistic population 
of axions is produced by flaton decay during the subsequent reheating,
whose density is constrained by nucleosynthesis. 
We compute all of the relevant reaction rates and evaluate
the nucleosynthesis constraint. We find that 
 the KSVZ model is practically
ruled out, while the
DFSZ model has a sizable allowed region of parameter space.

\end{abstract}


\thispagestyle{empty}

\end{titlepage}

\def\baselinestretch{1.1}

\section{Introduction}

With the discovery of the instantons it was realized that the pure gradient
topological term of the QCD Lagrangian 
$\theta_{QCD} g_S^2/32 \pi^2 F \tilde{F} $
can generate important physical consequences. In fact the induced CP 
violation  affects the electric dipole moment of the neutron suggesting 
the limit $\theta_{QCD}\leq 10^{-10}$.
The most attractive explanation for the origin of such a small parameter 
would be the Peccei-Quinn mechanism \cite{peccei}.
There is supposed to be a spontaneously broken global $U(1)$ symmetry 
(PQ symmetry), which is also explicitly broken by the color anomaly.
The corresponding pseudo-Goldstone boson is called the axion
\cite{kim}.

The PQ symmetry acts on some set of fields $\phi_i$ with charges
$Q_i$,
\be
\phi_i\to e^{iQ_i \alpha} \phi_i \,.
\label{pqsym}
\ee
To
generate the require $U(1)_{PQ}\times SU(3)_c\times SU(3)_c$ anomaly
we must choose appropriate particle spectra.
Depending on the charge of the SM matter content, we can have the
KSVZ (hadronic)  models \cite{ksvz} in which 
only some extra heavy quark fields are PQ charged
or the DFSZ \cite{dfsz} models in which, beyond the extra matter content
(at least two Higgs fields), also the Standard Model (SM)
matter is PQ charged.

Denoting the vacuum expectation values of the scalar fields charged under PQ
 by $v_i/\sqrt 2$, the PQ symmetry breaking scale $\vpq$ is 
defined as $\vpq^2=\sum_i Q_i^2v_i^2$, and the axion mass is given by
$\vpq m\sub a = (79\MeV)^2 N$ where $N$ is the number of quarks with
PQ charge.
Collider and astrophysics constraints require 
$\vpq\gsim 10^9\GeV$, 
and defining as usual $\mpl=(8\pi G_N)^{-1/2}
=2.4\times 10^{18}\GeV$ this allows the
range 
\be
10^9\GeV \lsim \vpq \lsim \mpl \,.
\ee
With typical assumptions about the cosmology, the requirement that 
axions give at most critical density places $\vpq$ towards the bottom of 
this range. In the particular case that the axions are radiated by 
strings with no subsequent entropy production,
one probably requires \cite{paul} $\vpq\sim 10^{10}\GeV$.

In a model with unbroken supersymmetry, the holomorphy of the 
superpotential ensures that PQ symmetry \eq{pqsym}
is accompanied by a symmetry acting on the radial parts of the
PQ charged fields
\be
\phi_i\to e^{Q_i \alpha} \phi_i \,.
\label{saxsym}
\ee
The corresponding pseudo-Goldstone boson is called the saxion
(or saxino), and the spin-half partner is called the axino.
Soft supersymmetry breaking gives the saxion a mass of order
$100\GeV$, and the axino typically has a mass of the same order
\cite{chunlukas} though it may be very light in special cases.\footnote
{We assume gravity-mediated 
supersymmetry breaking, which typically gives soft scalar masses in the range
$100$ to $1000\GeV$, and 
use the former estimate for definiteness.}

Let us consider the potential of the fields $\phi_i$
which break PQ symmetry. In a supersymmetric model there have to be at 
least two, but let us 
pretend for the moment that there is only 
one. Its potential will be 
of the form
\be
V=V_0-m^2 |\phi|^2+\frac14 \lambda\phi^4 +
\sum_{n=1}^\infty
 \lambda_n \frac{|\phi|^{2 n +4}}{\mpl^{2n}}
\,.
\label{phipot}
\ee
The non-renormalizable terms are expected to 
have coefficients $\lambda_n\sim 1$.
If the renormalizable coupling $\lambda$ is also of order
$1$, $m\sim \vpq$ and the non-renormalizable terms are negligible. 
(Remembering that there are at least two complex fields,
only the particular combination of fields corresponding
to the saxion will have the soft mass of order $100\GeV$.)
In this sort of model one can hope to understand a
value $\vpq\sim 10^{10}\GeV$ since that is the supersymmetry 
breaking scale \cite{kim84}, but it may be hard to understand a bigger 
scale.

We are concerned with a different class of models 
\cite{yamamoto,lps,cr,hitoshi1,hitoshi2}, in which $\phi$
represents a flat direction of supersymmetry. The quartic term is then
absent, while $m$ is a soft mass of order $100\GeV$.
If the $\phi^6$ term is present with unsuppressed coefficient
the vev is 
$\langle\phi\rangle\sim \lambda_1^{-1/4} 10^{10}$ GeV. If instead
the $\phi^8$ dominates one has
$\langle\phi\rangle \sim \lambda_2^{-1/6}10^{13}$ GeV  and so on.
For future reference, note that the mass of $|\phi|$ in the vacuum
is also of order of $m$,
and that the height of the potential is given by
\be
\(\frac{V_0^{1/4}}{10^6\GeV}\) \sim 
\(\frac{\langle\phi\rangle}{10^{10}\GeV} \)^{1/2} \,.
\ee
Fields of this kind, 
characterized by a large vev 
and a flat potential are called {\it flaton} fields, and the particles 
corresponding to them are called flatons \cite{thermal}. 

In a supersymmetric model 
there are $n\geq 2$ complex fields which all acquire vevs. 
We are interested in the case where these are flaton fields.
Then there are $2n-1$ flaton particles with mass of order 
$100\GeV$, and $n$ flatinos
with typically similar masses.
The saxion (axino) is a linear combination of the flaton
(flatino) fields, with no particular significance.

The rest of this paper is as follows.
In the next section we define our models and 
 summarize the cosmology.
In section \ref{2}, we give the  general
structure of the flaton  and flatino  masses.
In section \ref{3} we analyze 
the general self interactions between flatons and flatinos.
In section \ref{4} we 
see the effect of the interaction of the flatons with the matter fields
(The KSVZ flatons interact only with gluons  and gluinos  whereas 
DFSZ flatons interact also with ordinary matter and  supermatter.)
In section \ref{5} we find the parameter space regions that can satisfy
the cosmological constraints.
We conclude in Section \ref{6}.

\section{The model and its cosmology}

We consider a DFSZ and a KSVZ (hadronic) model. 

In both models, there 
are  two flaton fields $P$ and $Q$,
interacting with the superpotential \cite{hitoshi1,hitoshi2} 
\be
W\sub{flaton}=\frac{f}{\mpl}\hat{ P}^3 \hat{ Q} 
\,.
\ee
In the hadronic version, the interaction with matter
is 
\be \label{hadron}
W\sub{flaton-matter}=h_{E_i} \hat{ E}_i\hat{ E}^c_i\hat{ P}
\ee
where $E_i$ and $E_i^c$ are 
additional heavy quark and antiquark superfields.

In the DFSZ version, the interaction is
\be\label{peccei}
W\sub{flaton-matter}
=\frac{1}{2} \lambda \hat{ N}\hat{ N}\hat{ P}+\frac{g}{\mpl}  \hat{ H}_1 
\hat{ H}_2 
\hat{ P} \hat{ Q}
\ee
where $ \hat{N}$ are the right handed neutrino superfields and
$\hat{H}_{1,2}$ the two Higgs doublets.  Due to the second term we can
provide a solution to the $\mu$ problem \cite{nilles}.
In such  case we can add to the superpotential of 
the minimal supersymmetric standard model also the terms
$h_{\nu}\hat{l}\hat{H_2}\hat{N}$
that generate the necessary mixing between left and right neutrinos to
implement a see saw mechanism which can explain the solar 
and atmospheric neutrino deficits.

The cosmology of the DFSZ model has already been considered in a rough 
way \cite{eungjin}. Here we relatively complete treatment of both models.

To study the cosmology of the flaton fields, 
we can safety analyze only the superpotential 
$W\sub{flaton}$ for both 
models.
With the inclusion of the  soft
susy breaking terms, the potential is
\bear
V&=& m_P^2 |\phi\sub P|^2+
m_Q^2|\phi\sub Q|^2+
\frac{ f^2}{\mpl^2}  \(9  |\phi\sub P|^4 |\phi\sub Q|^2 +
|\phi\sub P|^6\)+
\(\frac{A_f}{\mpl} f \phi\sub P^3\phi\sub Q+h.c.\) \,.
\label{spot}
\eear
The soft parameters $m_P$, $m_Q$ and $A_f$ are all of order
$10^{2-3}\GeV$ in magnitude.
It is assumed that $m_P^2$ and $m_Q^2$ are both positive at the Planck 
scale. The unsuppressed interactions of $\phi_P$
 give radiative corrections
which drive  $m_P^2$ to a negative value  at the PQ scale,
triggering a vev for $\phi_P$.
As a result, 
when $\phi_P$  gets a vev, the tadpole term proportional to $A_f$ 
generates automatically a vev for the $\phi_Q$ field, both of them
are $v\sub Q\sim v\sub P
\sim 10^{10-12}\GeV$.

In the early Universe when $H\gsim 100\GeV$,
there will be effective values $m_P^2(t)$
and $m_Q^2(t)$. Supergravity interactions will make both of them
at least of order $H^2$ in magnitude.\footnote
{During inflation this result
might be avoided (say by $D$-term inflation) but it should still hold 
afterwards.} The cosmology depends on the signs of these effective 
values. If $m_P^2(t)$ is positive, 
$\phi_P$ is 
held at the origin by
the finite-temperature potential until
$T\sim |m\sub P|$. 
But the potential $V_0$ dominates the energy density in the 
regime $|m\sub P|\lsim T\lsim V_0^{1/4}$, leading to about
$\sim \ln\(V_0^{1/4}/|m\sub P|\)\sim$ 10 $e$-folds of 
thermal inflation \cite{thermal}.

To discuss what happens after thermal inflation, we suppose first that
$m_Q^2(t)$ is also positive in the early Universe, so that $\phi_Q$ 
is also trapped at the origin during thermal inflation.
When thermal inflation ends, $\phi_P$ moves away from the origin,
which destabilizes $\phi_Q$.
The fields $\phi_P$ and $\phi_Q$ move 
around an orbit
in field space, which would be closed if there were no energy loss.
If the only energy loss came from Hubble damping 
the fields would oscillate back and forth many times around
an almost-closed orbit. However, the parameter determining the strength 
of parametric resonance is $q\sim g\Phi_0/m\sim 10^8$, where $g\sim 1$ 
is a typical coupling, $\Phi_0\sim 10^{10}\GeV$ is the amplitude of the 
oscillation and $m\sim 10^2 \GeV$ is its angular frequency. One therefore
 expects 
that parametric resonance will efficiently damp the orbit, converting 
most of the energy into flaton particles. At first these particles are
marginally relativistic, but after a few Hubble times they become
non-relativistic. The rest of the energy resides in the homogeneous
oscillating flaton fields, now with small amplitude and therefore almost
simple harmonic motion corresponding to some more
non-relativistic flatons. The flatons decay leading to final reheating
at a temperature \cite{thermal}
\bea
T\sub {RH} &\simeq & 1.2 g_{RH}^{-\frac{1}{4}} \sqrt{\mpl \Gamma_{\phi}} 
      \nonumber
\sim  3\(\frac{10^{11}\GeV}{\vpq} \) \( \frac{m}{300\GeV} \)^3
\GeV \,.
\eea
where $m$ is the mass of the lightest flaton
and $g_{RH}\sim 100$ is the effective number of relativistic species
at reheat. 

Two different axion populations are produced.
One population is radiated by the PQ strings that form after thermal
inflation. They become dark matter with abundance \cite{paul}
$\Omega\sub a\sim (\vpq/10^{10}\GeV)^{1.2}$. The present scenario
{\em predicts
$\vpq\sim 10^{10}\GeV$, making the axion
an excellent dark matter candidate.} Note that 
in contrast with the 
general case, the axion density in this scenario cannot be reduced by
entropy production after the epoch $T\sim 1\GeV$
when the axions acquire mass; pre-existing long-lived particles that 
might do the job have been diluted away by the thermal inflation.

The other axion population, that is our main concern,
comes from the decay of the flatons \cite{thermal,chunlukas}. 
This 
population is still relativistic at nucleosynthesis, and its
density must satisfy the constraint
\be
\(\frac{\rho_a}{\rho_{\nu}}\)_{NT}\leq \delta N_{\nu}\sim 0.1-1.5
\,,\label{neutrino}
\ee
where $\rho_{\nu}$
 denotes the energy density of a single species of relativistic neutrino
and $\delta N_{\nu}$ the number of extra neutrino species allowed by
nucleosynthesis. 

If there were just one species of flaton field $\phi$, this would give the 
bound \cite{eungjin}
\bea
\(\frac{\rho_a}{\rho_{\nu}}\)_{NT} &= &
\frac{43}{7}
\(\frac{43/4}{g_{RH}}\)^{1/3} \frac{B_a}{1-B_a} \nonumber\\
&\simeq &
\frac{43}{7}
\(\frac{43/4}{g_{RH}}\)^{1/3} B_a \,,
\label{oneflat}
\eea
where
$B_a=\Gamma_a\(\phi\rightarrow a+a\)/\Gamma_{tot}$
is the branching ratio, and in the
last line we assumed $B_a\ll 1$.
{}From \eq{neutrino} we get the bound
\be
\frac{\Gamma\(\phi\rightarrow a+a\)}{\Gamma(\phi\rightarrow X)}
\leq 0.24 \(\frac{\delta N}{1.5}\)\(\frac{g_{RH}}{43/4}\)^{1/3} \,.
\label{onephibound}
\ee
Varying $T_{RH}$ from 6 MeV to $m\sim 100\GeV$ 
we get a factor two variation coming
 from the number of degrees of freedom.
For $\delta N_{\nu}$ varying from 0.1 to 1.5 we get 
$B_a<1/3$ to $B_a<0.02$.

Our model has three flaton particles 
and the quantity $B_a$ to be used in \eq{oneflat} is 
\bear          \nonumber
B_a &=& \sum_I r_I B_I \quad {\rm where}  \quad
         r_I= {n_I \over \sum_J n_J} \,, \\
 B_I &\equiv& {\Gamma\(I \to a+a\) + {1\over2} \Gamma\(I \to a+X\)
 \over \Gamma_{\rm tot}\(I\) } \,.
\eear
Here $n_I$
is the number density of the $I$th flaton just before reheating, 
related to the mass density $\rho_I$ by
$n_I=\rho_I/m_I$. In principle one could calculate the $n_I$ 
(coming from parametric resonance and some residual homogeneous
flaton oscillations
decay) but we have not done that, and to estimate the allowed parameter 
range for the model we shall simply assume $r_I=1/3$.

Next consider what happens after thermal inflation in the case that
$m_Q^2(t)$ is negative.
During thermal inflation, while $\phi_P$ is trapped at the origin, 
$|\phi_Q|$ will have some value $\gg\vpq$ determined by a
higher-order non-renormalizable term.
At the end of thermal 
inflation, all three flaton particles will be produced in the manner we 
have described, and the nucleosynthesis constraint 
still holds. The difference from the previous case is that 
axionic strings are not produced, so that dark matter axions are 
produced only by the quantum fluctuation of the axion field during 
inflation. The density is now \cite{turner,myaxion}
$\Omega\sub a\sim 3(\theta/\pi)^2 (\vpq/10^{12}\GeV)^{1.2}$,
where $\theta<\pi$ is the misalignment angle. Again, thermal inflation 
means that the abundance will not be diluted by entropy production,
so discounting an accidentally small $\theta$ we again have a rather 
definite prediction, which is too low. For the axions to be the dark 
matter in this scenario, one would have to increase $\vpq$ by generating 
it from a higher-order non-renormalizable term.

Finally, if
$m_P^2(t)$ is negative in the early Universe,
there is no thermal inflation and one is back with all the uncertainties 
of more general models \cite{myaxion,lyst,myaxion2}. 
Axion cosmology now depends on the scale of the 
inflaton potential, the reheat temperature after inflation, the
decoupling temperature of the flatons and so on. We have nothing to say 
about that case.

\section{Flaton and flatino spectrum}
\label{2}

\subsection{Flaton spectrum}

We write the flaton fields as
\bea
\phi\sub P &=& \frac{v_P+P}{\sqrt{2}}  e^{i \frac{A_P}{v_P}} \nonumber\\
\phi\sub Q &=& \frac{v_Q+Q}{\sqrt{2}}  e^{i \frac{A_Q}{v_Q}} \,,
\eea
and we shall take $v_P$, $v_Q$, $A_f$ and $f$ as the independent 
parameters in the potential \eq{spot}.
The main components of the axion field are
\be
a= -\frac{v_P}{F_{PQ}} A_P +3 \frac{v_Q}{F_{PQ}} A_Q 
\ee
where $F_{\rm PQ}^2=v_P^2+9v_Q^2$  and we
 have neglected the $O\(v_{EW}/F_{\rm PQ}\)$
components along the $H_{1,2}$ directions.
The orthogonal field to the axion (both are CP Odd)  corresponds to a 
flaton particle. It is
\be
\psi'=
-\frac{v_P}{F_{\rm PQ}} A_Q -
3 \frac{v_Q}{F_{\rm PQ}} A_P
\ee
which has a mass
\be
M_{\psi'}^2=-\frac{f A_f v_P F_{\rm PQ}^2}{2  \mpl v_Q}
=-\frac{f}{g} \mu A_f \(x^2+9\) \,.
\ee
where 
\bea
\frac{\mu}{g} &\equiv& \frac{ v_P v_Q}{2 \mpl} 
\label{gmu} \\
x &\equiv & \frac{v_P}{v_Q} \,.
\eea
For future convenience we have introduced a
 quantity $\mu$, related to the $g$ appearing 
only in the DFSZ model. At this stage 
 results
depend only on the ratio $\mu/g$ defined by (\ref{gmu})
 and they apply to both models.
Since $M^2_{\psi'}$ is positive,
 $A_f$ and $f$ must have opposite
signs.

The other two flaton particles correspond to the
CP even fields $P$ and $Q$. They have a 2$\otimes$2 mass matrix whose
components are 
\bear
M^2_{QQ}&=& M_{\psi'}^2 \frac{x^2}{9+x^2} \\\nonumber
M^2_{PQ}&=& 9  f^2 \frac{ v_P^4}{ \mpl^2 x}- 
3\frac{ M_{\psi'}^2 x}{ 9+x^2}=
3 x\(12  \frac{f^2}{g^2} \mu^2  - 
 \frac{ M_{\psi'}^2 }{ 9+x^2}\)
 \\\nonumber
M^2_{PP}&=&3  f^2 \frac{v_P^4\(x^2+3 \)}{ \mpl^2 x^2 }-
3 \frac{M_{\psi'}^2 }{9+x^2}=
12  \frac{f^2}{g^2} \(x^2+3 \) \mu^2 -
3 \frac{M_{\psi'}^2 }{9+x^2} \,.
\\\nonumber
\eear
Here two mass parameters $m_P^2, m_Q^2$ in Eq.~(\ref{spot}) are replaced 
in favor of $v_P, v_Q$.
Performing the rotation from the flavor basis $||P\; Q||$ to the mass basis
$||F_1\; F_2||$
\bea
 P &=& \cos\alpha F_2 -\sin\alpha F_1 \nonumber \\
 Q &=& \sin\alpha F_2 + \cos\alpha F_1  \,,
\eea
the mixing angle results
\be \label{mixing}
\tan 2 \alpha=\frac{2 M_{PQ}^2}{M_{QQ}^2-M_{PP}^2}=-6 \frac{x}{x^2+3}\equiv -a 
\,.
\ee
Note that 
$\tan 2 \alpha$
ranges from  $ -\sqrt{3}$  for $x=\sqrt{3}$
to zero for  $\;\;x\gg1$  and $x\rightarrow 0$.
For each $x$ there are two solutions given by
\bea
\cos^2 \alpha_1 &=& \frac{1}{2} \(1-\frac{1}{\sqrt{1+a^2}}\) \\
\cos^2 \alpha_2 &=& \frac{1}{2} \(1+\frac{1}{\sqrt{1+a^2}}\)
\eea
with $\sin \alpha_{1,2}\geq 0$.
The two solutions will be relevant when we discuss
the   parameter space for the decay of flatons into axions.

The two eigenstates have masses

\be
M^2_{F_{2,1}}=\frac{\mu^2}{2}\left(
\frac{f}{g}(12(x^2+3) \frac{f}{g}+
(3-x^2)\frac{A_f}{\mu})\pm
|\frac{f}{g} (12 \frac{f}{g}+
\frac{A_f}{\mu}) |\sqrt{
9+42 x^2 +x^4}\right)
\ee

The requirement of a positive definite
spectrum ($m^2_{F_1}>0)$ gives the constraint  
\be
y_1<y=-\frac{g \;A_f}{f\; \mu}
\frac{9+x^2}{ 4 x^2}<y_2 \,,
\ee
where
\be 
y_{1,2}\equiv \frac{9+x^2}{ 8 x^2}\(21+x^2 \pm 
\sqrt{9+42 x^2+x^4}\)
\ee
or
\be \label{f1bound}
 \frac{1}{2}\(21+x^2 -
\sqrt{9+42 x^2+x^4}\)<-\frac{g \;A_f}{f\; \mu}<\frac{1}{2}\(21+x^2 +
\sqrt{9+42 x^2+x^4}\)
\ee
and requiring positive diagonal elements implies  also
\be
y< y_3 \equiv \frac{\(x^2+3\)\(x^2+9\)}{x^2}
\ee

\subsection{Flatino spectrum  }

{}From the   superpotential $W\sub{flaton}$
we can directly extract   also the flatino's mass matrix
whose eigenvalues are
\be
M^2_{\tilde{F}_{2,1}}=\frac{9}{4} \frac{M^2_{\psi'}}{y\; x^2}[x^2 +2 \pm 
 \sqrt{x^2+1}]=9\frac{f^2}{g^2}\mu^2[x^2 +2 \pm 
 \sqrt{x^2+1}] 
\ee
The eigenstates $\tilde{F}_1, \tilde{F}_2$ are related to the 
flavor states $\tilde{P}, \tilde{Q}$ by 
\bear
\tilde{F}_1 &=& \cos \tilde{\alpha} \tilde{P}+ \sin  \tilde{\alpha}
\tilde{Q}  \nonumber\\
\tilde{F}_2 &=& -\sin  \tilde{\alpha}
 \tilde{P} +  \cos \tilde{\alpha} \tilde{Q} 
\label{angle}
\eear
where $\tan 2 \tilde{\alpha}=x$,


A parameter space analysis indicates that  we have always 
$M_{F_1} \geq  2 M_{\tilde{F}_{1}}$.
 This automatically  forbids  the decay of $F_1$ to 
flatinos
leaving  open only the decay into flatinos of the heavier 
 $F_2$ and  $\psi'$  flatons.

\section{Interactions  between flatons  }
\label{3}

Now that we know the general mass matrix structure of 
the flatons and flatinos, common 
  both to  the DFSZ and KSVZ models, we can start 
analyzing   the  various decay rate between  flatonic fields.\footnote
{The two body decay rate is given by the expression
\be
\Gamma \(i\rightarrow f_1 f_2\)= 
\frac{1}{16 \pi}
\frac{S |P|}{M_i^2} \int^1_{-1} d \cos \theta A\(E_1=E_2, \cos \theta\)
\ee
where S=1 if $f_1 \neq f_2$ and $S=1/2$ is $ f_1=f_2$ ;
$E_1=E_2= \(M_i^2-M_{f_1}^2-M_{f_2}^2\)/2 M_i$,
$ |P|= 1/\(2 M_i\) \sqrt{\(M_i^2-\(M_{f_1}+M_{f_2}\)^2\)
\(M_i^2-\(M_{f_1}-M_{f_2}\)^2\)}$ and 
$A\(..\)$  is the square amplitude of the decay.}
%
%
We begin with the decay channels induced by 
the kinetic term and 
the superpotential $W\sub{flaton}$, which are common to
the KSVZ and the DFSZ models.

\subsection{ Derivative  and cubic interaction terms between flatons}

The flaton interaction terms with at least one derivative are given by the 
lagrangian 
\bear 
L_{\partial}&=&
\frac{2\, v_P \, P+P^2}{2 v_P^2} [
\frac{v_P^2}{F_{\rm PQ}^2} \(\partial a\)^2
+9\frac{v_Q^2}{F_{\rm PQ}^2} \(\partial \psi'\)^2 +
6 \frac{v_Q v_P}{F_{\rm PQ}^2} \partial \psi' \partial a]+ \nonumber\\
&&\frac{2\, v_Q \, Q+Q^2}{2 v_Q^2} [ 
9 \frac{v_Q^2}{F_{\rm PQ}^2} \(\partial a\)^2
+\frac{v_P^2}{F_{\rm PQ}^2} \(\partial \psi'\)^2 -
6 \frac{v_Q v_P}{F_{\rm PQ}^2} \partial \psi' \partial a]
\label{derivas}\eear
{}From this expression we can extract the following terms expressed 
in mass eigenstates.

The trilinear derivative interactions with no axions
\be
\(\partial \psi'\)^2\frac{1}{F_{\rm PQ} x \sqrt{x^2+9}}
[\(-9\sin \alpha +x^3 \cos \alpha\) F_1 +
\(9 \cos \alpha +x^3 \sin \alpha\)F_2] \,,
\label{psis}
\ee
The  trilinear derivative interactions with only one axion
\be
L_{ F_i \psi'a}=  \partial \psi' \partial a  \frac{6} 
{F_{\rm PQ}\, \sqrt{x^2+9}}[
 \(\cos \alpha -x \sin \alpha\)F_2-\(\sin \alpha + x \cos \alpha\)F_1] \,,
\ee
The  trilinear derivative interactions with  two axions
\be
L_{F_i a a}  =
|\partial_{\mu}a|^2   \frac{1}{F_{\rm PQ} \sqrt{x^2+9}}\( \(9 \cos \alpha-x\,
 \sin \alpha\)F_1+ \(9\, \sin \alpha +x\, \cos \alpha\) F_2\) \,.
\ee
All the above derivative interactions can be 
transformed in scalar interactions if  we are working at tree level and with
on-shell  external particles
\be
\phi\sub 1 \(\partial_{\mu} \phi\sub 2\) \(\partial^{\mu} \phi\sub 3\)=
\frac{1}{2} \(M_{\phi\sub 1 
}^2-M_{\phi\sub 2}^2-M_{\phi\sub 3}^2\)\phi\sub 1\phi\sub 2\phi\sub 3
\ee

The cubic interactions come also from the flatonic 
superpotential and the soft terms
\begin{eqnarray} 
L_{\phi^3}&=&\({9\over2}f^2
 {v_P v_Q^2 \over \mpl^2} +{5\over2}f^2{v_P^3 \over
 \mpl^2} -
    \frac{ M_{\psi'}^2 v_Q x}{v_P^2 \(x^2+9\)} \) P^3   
+ \({27\over2}f^2{v_P^2 v_Q\over \mpl^2}
-3\frac{ M_{\psi'}^2  x}{v_P \(x^2+9\)} \)P^2Q
\nonumber \\
&+& \({9\over2}f^2{v_P^3\over \mpl^2}\) PQ^2
+ \({3f\over4}{A_fF_{\rm PQ}^2 \over \mpl v_Q}\)P\psi'\psi'
+ \({f\over4}{A_fF_{\rm PQ}^2v_P \over \mpl v_Q^2}\)Q\psi'\psi' \,.
\label{cubis}\end{eqnarray}
In the mass basis, we get
\bear L_{F_2F_1F_1}&=&\nonumber
\left\{ -3\frac{M_{\psi'}^2\,\sin \alpha}
{F_{\rm PQ}x \sqrt{9 + x^2}} 
\,\( -2 x\cos^2 \alpha + 
        x\,\sin^2\alpha +
        \,\cos\alpha\,\sin\alpha 
 \) + \right.
 \\\nonumber 
 &&
 \frac{6\;f^2\,\mu^2 \sqrt{x^2+9}}
{g^2 x F_{\rm PQ}} \( -18\,\cos^2\alpha\,\sin\alpha \,x +
     9\,\sin^3\alpha\,x + 3 \,\cos^3\alpha \,x^2 \right. \\
&& \left.\left. -\cos\alpha\, \sin^2\alpha (-9 + x^2) \)    \right\}
F_1^2 F_2 \equiv \frac{A_{F_2F_1F_1}}{2}F_1^2 F_2
\eear
For the  full trilinear $F_2 \psi'\psi'$ interaction, we have to add up
the terms in Eqs.~(\ref{psis}) and (\ref{cubis}) to obtain
%
%
\bear
L_{F_2 \psi'\psi'}=
\frac{ M_{F_2}^2 }{2 x \sqrt{9+x^2}F_{\rm PQ}}
\(-\(3 \sin \alpha +x \cos \alpha\)\(x^2+9\)
\frac{M_{\psi'}^2}{M_{F_2}^2}+\right.
\\\nonumber
\left.
\(9 \cos \alpha +\sin \alpha x^3\)\(1-2\frac{M_{\psi'}^2}{M_{F_2}^2}\)\)
F_2 \psi'\psi'\equiv \frac{A_{F_2 \psi'\psi' }}{2}\psi'^2 F_2
\eear
%

Collecting the above formulae one finds the  decay rates 
among flatons and axions  

\begin{eqnarray}
&&\Gamma\(F_2\to aa\) = {1\over 32\pi} {M^3_{F_2}\over F^2_{\rm PQ} \(x^2+9\) }
                      \(x\cos\alpha + 9 \sin\alpha\)^2 \nonumber\\
&&\Gamma\(F_1\to aa\) = {1\over 32\pi} {M^3_{F_1}\over F^2_{\rm PQ} \(x^2+9\) }
                      \(-x\sin\alpha + 9 \cos\alpha\)^2 \nonumber\\
&&\Gamma\(F_2\to F_1F_1\) =
\frac{1}{32 \pi M_{F_2} }\sqrt{1-4 \frac{M^2_{F_1}}{M^2_{F_2}}}
|A_{F_2F_1F_1}|^2 \nonumber\\
&&\Gamma\(F_2\to \psi'\psi'\) = 
\frac{1}{32 \pi M_{F_2} }\sqrt{1-4 \frac{M^2_{\psi'}}{M^2_{F_2}}}
|A_{F_2\psi'\psi'}|^2 \\
&&\Gamma\(F_2\to a\psi'\) = {1\over 16\pi} {M^3_{F_2}\over F^2_{\rm PQ} 
   \(x^2+9\) } \(1-{M^2_{\psi'} \over M^2_{F_2}}\)^{3}
              \(3\cos\alpha-3x\sin\alpha\)^2 \nonumber\\ 
&&\Gamma\(\psi'\to aF_2\) = \Gamma\(F_2\to a\psi'\)
    \left|_{M^2_{\psi'} \leftrightarrow M^2_{F_2} }  \right. \nonumber\\
&&\Gamma\(\psi'\to aF_1\) = \Gamma\(F_1\to a\psi'\)
    \left|_{M^2_{\psi'} \leftrightarrow M^2_{F_1} }  \right.  \nonumber
\end{eqnarray}

Energy conservation will of course forbid some of these reactions,
depending on the flaton masses. 
As  $M_{F_1}<M_{\psi'}$ 
the channels $F_1\rightarrow \psi'\psi'$ and
$F_1\rightarrow \psi'a$ are always forbidden.

\subsection{  Interaction terms between flatons  and flatinos }

The trilinear Lagrangian terms responsible for the decay of flatons or 
flatinos are   
\bear 
L_{\phi  \bar{\tilde{\phi}} \tilde{\phi}}&=&
\frac{3 f}{2\mpl}\(\(v_PQ+v_QP\)\bar{ \tilde{P}} \tilde{P}+
i \frac{v_P^2+3 v_Q^2}{F_{\rm PQ}} \psi' \bar{
\tilde{P}}\gamma_5 \tilde{P}
-i2 \frac{v_P v_Q}{F_{\rm PQ}} a \bar{\tilde{P}}\gamma_5 \tilde{P}\)
+ \nonumber\\
&& \frac{3 f}{2 \mpl}\(2 P v_P\bar{ \tilde{P}} \tilde{Q} +
2i \frac{v_P}{F_{\rm PQ}} \(3 v_Q \psi'+v_P a\)\bar{\tilde{P}}\gamma_5\tilde{Q}
\)
\label{yukas} \eear
(the tilded fields are the fermionic
 superpartner of the respective $P$ and $Q$ scalars).
Let us denote the Yukawa couplings between  the flaton (or the axion) and 
the flatinos in mass basis 
by $-L_{Yuk}=Y_{ijk} \phi_i \bar{\tilde{F}}_j\(1,\gamma_5\)
\tilde{F}_k/2$ where $\gamma_5$ is taken for $\phi_i=a, \psi'$.
We find from Eq.~(\ref{yukas}) the following expressions for the
Yukawa couplings 
\bear
&& Y_{F_1\tilde{F}_1\tilde{F}_1}= {6f\mu\sqrt{x^2+9} \over gxF_{\rm PQ} }
[\(x\cos\alpha-\sin\alpha\)\cos^2\tilde{\alpha}-x\sin\alpha\sin2\tilde{\alpha}]
  \nonumber \\
&& Y_{F_1\tilde{F}_2\tilde{F}_2}= {6f\mu\sqrt{x^2+9} \over gxF_{\rm PQ} }
[\(x\cos\alpha-\sin\alpha\)\sin^2\tilde{\alpha}+x\sin\alpha\sin2\tilde{\alpha}]
  \nonumber \\
&& Y_{F_1\tilde{F}_1\tilde{F}_2}= {6f\mu\sqrt{x^2+9} \over gxF_{\rm PQ} }
[-x\sin\alpha\cos2\tilde{\alpha}+{1\over2}\(\sin\alpha-x\cos\alpha\)
  \sin2\tilde{\alpha}]      \nonumber \\
&& Y_{F_2\tilde{F}_1\tilde{F}_1}= {6f\mu\sqrt{x^2+9} \over gxF_{\rm PQ} }
[\(x\sin\alpha+\cos\alpha\)\cos^2\tilde{\alpha}+x\cos\alpha\sin2\tilde{\alpha}]
  \nonumber \\
&& Y_{F_2\tilde{F}_2\tilde{F}_2}= {6f\mu\sqrt{x^2+9} \over gxF_{\rm PQ} }
[\(x\sin\alpha+\cos\alpha\)\sin^2\tilde{\alpha}-x\cos\alpha\sin2\tilde{\alpha}]
  \nonumber \\
&& Y_{F_2\tilde{F}_1\tilde{F}_2}= {6f\mu\sqrt{x^2+9} \over gxF_{\rm PQ} }
[+x\cos\alpha\cos2\tilde{\alpha}-{1\over2}\(\cos\alpha+x\sin\alpha\)
  \sin2\tilde{\alpha}]      \nonumber \\
&& Y_{\psi'\tilde{F}_1\tilde{F}_1}= {6f\mu \over gxF_{\rm PQ} }
[-\(3+x^2\)\cos^2\tilde{\alpha}-3x\sin2\tilde{\alpha}]           \\
&& Y_{\psi'\tilde{F}_2\tilde{F}_2}= {6f\mu \over gxF_{\rm PQ} }
[-\(3+x^2\)\sin^2\tilde{\alpha}+3x\sin2\tilde{\alpha}] \nonumber \\
&& Y_{\psi'\tilde{F}_1\tilde{F}_2}= {6f\mu \over gxF_{\rm PQ} }
[-3x\cos2\tilde{\alpha}+{1\over2}\(3+x^2\)\sin2\tilde{\alpha}] \nonumber \\
&& Y_{a\tilde{F}_1\tilde{F}_1}= {6f\mu \over gxF_{\rm PQ} }
[2x\cos^2\tilde{\alpha}-x^2\sin2\tilde{\alpha}] \nonumber \\
&& Y_{a\tilde{F}_2\tilde{F}_2}= {6f\mu \over gxF_{\rm PQ} }
[2x\sin^2\tilde{\alpha}+x^2\sin2\tilde{\alpha}] \nonumber \\
&& Y_{a\tilde{F}_1\tilde{F}_2}= {6f\mu \over gxF_{\rm PQ} }
[-x^2\cos2\tilde{\alpha}-x\sin2\tilde{\alpha}] \nonumber 
\eear
%
%
%
{}From this we can extract the decay rates for
$F_i\to \tilde{F_j}\tilde{F_k}$,
$\psi'\to \tilde{F_j}\tilde{F_k}$, or
$\tilde{F_2} \to \tilde{F_1} F_i \(\psi',a\)$ 
\begin{eqnarray}
&&\Gamma\(F_i\to \tilde{F_j}\tilde{F_k}\) =
   {M_{F_i} \over 8\pi } S
   \( 1-{(M_{\tilde{F_j}}+M_{\tilde{F_k}})^2\over M^2_{F_i}} \)^{3\over2}
   \( 1-{(M_{\tilde{F_j}}-M_{\tilde{F_k}})^2\over M^2_{F_i}} \)^{1\over2}
   Y^2_{F_i\tilde{F_j}\tilde{F_k}}  \nonumber\\
&&\Gamma\(\psi'\to \tilde{F_j}\tilde{F_k}\) =
   {M_{\psi'} \over 8\pi } S 
   \(1-{(M_{\tilde{F_j}}+M_{\tilde{F_k}})^2\over M^2_{\psi'}} \)^{1\over2}
   \(1-{(M_{\tilde{F_j}}-M_{\tilde{F_k}})^2\over M^2_{\psi'}} \)^{3\over2}
   Y^2_{\psi'\tilde{F_j}\tilde{F_k}}  \\
&&\Gamma\(\tilde{F_2}\to \tilde{F_1}F_i\) =
   {M_{\tilde{F_2}} \over 16\pi } 
   \(1-{(M_{\tilde{F_1}}+M_{{F_i}})^2\over M^2_{\tilde{F_2}}}\)^{1\over2}
   \(1-{(M_{\tilde{F_1}}-M_{{F_i}})^2\over M^2_{\tilde{F_2}}}\)^{1\over2}
   \( \(1+{M_{\tilde{F_1}} \over M_{\tilde{F_2}}}\)^2 -
          {M^2_{F_i} \over M^2_{\tilde{F_2}}} \)
   Y^2_{F_i\tilde{F_1}\tilde{F_2}}  \nonumber\\
&&\Gamma\(\tilde{F_2}\to \tilde{F_1}\psi'\) =
   {M_{\tilde{F_2}} \over 16\pi } 
   \(1-{(M_{\tilde{F_1}}+M_{\psi'})^2\over M^2_{\tilde{F_2}}}\)^{1\over2}
   \(1-{(M_{\tilde{F_1}}-M_{\psi'})^2\over M^2_{\tilde{F_2}}}\)^{1\over2}
   \( \(1-{M_{\tilde{F_1}} \over M_{\tilde{F_2}}}\)^2 -
          {M^2_{\psi'} \over M^2_{\tilde{F_2}}} \)
   Y^2_{\psi'\tilde{F_1}\tilde{F_2}}  \nonumber \\
&&\Gamma\(\tilde{F_2}\to \tilde{F_1}a\) = 
   {M_{\tilde{F_2}} \over 16\pi } 
   \(1-{M_{\tilde{F_1}}^2\over M^2_{\tilde{F_2}}}\)
   \( 1-{M_{\tilde{F_1}} \over M_{\tilde{F_2}}}\)^2 
   Y^2_{a\tilde{F_1}\tilde{F_2}}  \nonumber
\end{eqnarray}
where $S$ is a symmetric factor ($1/2$ for identical final states or otherwise
 $1$).

\section{Interaction of flatons and flatinos with matter fields}
\label{4}

Now we study the interactions of the flatons with 
 matter  and supermatter, specified by
\eq{hadron} for the KSVZ case and by \eq{peccei} for the DSFZ case.
Through these interactions the over-produced flatons
or flatinos could decay into ordinary matter while the number 
of the decay produced axions are sufficiently suppressed satisfying the
nucleosynthesis limit (\ref{neutrino}).

\subsection{KSVZ model: Interactions between flatons  and gluons }

We begin with the hadronic models  in which the only decay mode available
for the flatons is into two gluons coming from the anomaly 
(when the space phase will be available, we have to take into account  also
the decay into  massive gluinos, 
in this discussion we neglect such a possibility).
The respective one loop corrected  decay rates are
\be
\Gamma \(F_1 \rightarrow g+g\)=  \frac{\alpha_S^2\(M_{F_1}\)}{ 72 \pi^3}\,
N^2_E \, \frac{M_{F_1}^3}{x^2\;F_{\rm PQ}^2}\(x^2+9\) \sin^2\alpha
\(1+\frac{95}{4}\frac{\alpha_S\(M_{F_1}\)}{ \pi}\)
\ee
and
\be
\Gamma \(F_2 \rightarrow g+g\)=  \frac{\alpha_S^2\(M_{F_2}\)}{ 72 \pi^3}\,N^2_E\,
\frac{M_{F_2}^3}     {x^2\;F_{\rm PQ}^2}\(x^2+9\) \cos^2\alpha
\(1+\frac{95}{4}\frac{\alpha_S\(M_{F_2}\)}{ \pi}\)
\ee
where $N_E$ is the total number of the
 superheavy exotic quark fields ($M_E=h_E v_P \gg M_{F_i}$).

We do not consider the flatino decay into a gluon and a gluino which
will be irrelevant for our discussion.

\subsection{DFSZ model:
 Interactions between flatons/flatinos and ordinary matter}

The decay properties of the flatons in the DFSZ models involves the 
direct interactions between flatons and ordinary matter 
and supermatter 
In general the interaction between flatons and Higgs fields
are quite interesting  due to the fact that this two sectors,
after the spontaneous breaking of the PQ and the EW symmetry, mix together.
We notice that the Peccei-Quinn symmetry prevents the introduction of a  
SUSY invariant mass term $\mu H_1H_2$,  solving automatically   
the so called $\mu$ mass problem as mentioned before.

Let us start by writing the Higgs-flaton potential
\begin{eqnarray} 
V\(H,\phi\) &=&
 \left| H_1 \right|^2
\( m_{H_1}^2 + \left| g{\phi\sub P \phi\sub Q \over \mpl} \right|^2 \)
+ \left| H_2 \right|^2
\( m_{H_2}^2 + \left| g{\phi\sub P \phi\sub Q \over \mpl} \right|^2 \)
 \nonumber  \\
 &&
+ \left\{ g H_1 H_2 \( A_g{\phi\sub P\phi\sub Q \over \mpl} +
   3f^*{\phi\sub P^{*2}|\phi\sub Q|^2 \over \mpl^2} +
   f^*{\phi\sub P^{*2}|\phi\sub P|^2 \over \mpl^2} \) +
 \mbox{c.c.} \right\} \\
&&
 + {1\over8} \(g^2+ {g'}^2\) \( |H_1|^2 - |H_2|^2 \)^2  \nonumber \,.
\label{VHf}
\end{eqnarray}
When the fields $\phi\sub {P,Q}$ get  vevs, the  
$m_3^2H_1H_2$ mass term is generated dynamically.
The size of such a term is fixed  by
\be
m_3^2=
\mu\(A_g  +\frac{f}{g} \mu \(x^2+3\)\)
\ee
In the limit  $|m_3^2| \gg M_W^2$ the masses of the pseudoscalar $A^0$, of
the CP even scalar Higgs field $H^0$
and of the charged Higgs fields $H^{\pm}$
are almost degenerate
\be
m_{A^0,H^0,H^{\pm}}^2 \simeq -\frac{ m_3^2}{\sin \beta \cos \beta}
\ee
so from  the constraint of positivity of such a masses we get
\be \label{pseudo}
\frac{A_g}{\mu}+\frac{f}{g}(x^2+3) \leq 0
\ee

%
%
In such a limit we also know that the mass eigenstate of the CP even 
electroweak sector $H^0,h^0$ and of the CP odd one $A^0,G^0$  are
\bear
H^0&=& -\sin \beta \;  h_1^0 +\cos \beta\; h_2^0 \nonumber \\
h^0&=& \cos \beta \;h_1^0 +\sin \beta \;h_2^0 \\
A^0&=& \sin \beta \;A_1^0 +\cos \beta \;A_2^0  \nonumber\\
G^0&=& \cos \beta \;A_1^0 -\sin \beta \;A_2^0  \nonumber 
\eear
where $H_1=\frac{1}{\sqrt{2}}\( v_1+h_1^0+iA_1^0\)$ and 
$ H_2= \frac{1}{\sqrt{2}}\( v_2+h_2^0+iA_2^0\)$  are the gauge eigenstates
and $\tan\beta=v_2/v_1$.
To allow the flaton decay into $A^0$,
we want it to be light so that small $\tan\beta$ is preferred in our discussion.
Hereafter we will take $\tan\beta=1$.

{}From Eq.~(\ref{VHf}),  we find 
\begin{eqnarray}
&&V_{Fhh}={1\over2} \mu^2 \(h_1^{02}+h_2^{02}+ A_1^{02}+A_2^{02}\)
                \({P \over v_P}+ {Q\over v_Q}\)
+  {1\over2} \left[\(h_1^0h_2^0-A_1^0A_2^0\) +
                    i \(h_1^0A_2^0+ h_2^0A_1^0\)\right] \\
&&\left[ A_g\mu\({P\over v_P} + {Q \over v_Q} - i {x^2+3 \over x F_{\rm PQ}} 
   \psi'\) + 6 {f\over g} \mu^2 \({P \over v_P} + {Q \over v_Q} + 
  i{3\over xF_{\rm PQ}}\psi'\)
  + x^2{f\over g} \mu^2 \( 4{P \over v_P} + i{6\over xF_{\rm PQ}}\psi'\) \right]
  \nonumber\\
  && + c.c. \nonumber
\end{eqnarray}
It is then simple manner to get  the decay rates for 
the kinematically  more favorable decay channels
 $F_{1,2} \rightarrow h^0h^0$ and $\psi \to h^0 A^0$
\begin{eqnarray} 
&&\Gamma\(F_1\to h^0h^0\)=  
   {M^3_{F_1} \over 32\pi F_{\rm PQ}^2} \frac{\(x^2+9\)}{16\, x^2}
 {\mu^4\over M^4_{F_1}}
   \(1-{4M_{h^0}^2\over M_{F_1}^2}\)^{1/2} |A_{F_1 h h}|^2  \nonumber \\
&&\Gamma\(F_2\to h^0h^0\)=
   {M^3_{F_2} \over 32\pi F_{\rm PQ}^2}  \frac{\(x^2+9\)}{16\, x^2} 
{\mu^4\over M^4_{F_2}}
   \(1-{4M_{h^0}^2\over M_{F_2}^2}\)^{1/2}  |A_{F_2 h h}|^2 \\
&&\Gamma\(\psi'\to h^0A^0\)=  
  {M^3_{\psi'} \over 16\pi F_{\rm PQ}^2} {\mu^4\over M^4_{\psi'} }
\(1-{\(M_{h^0}-M_{A^0}\)^2\over M_{\psi'}^2}\)^{1/2}
  \(1-{\(M_{h^0}+M_{A^0}\)^2\over M_{\psi'}^2}\)^{1/2}
|A_{\psi h A}|^2 \nonumber
\label{Fhh}
\end{eqnarray}
where 
\begin{eqnarray}
&& A_{F_1 hh} =\sin2\beta [ \({A_g\over \mu}
        + 6{f\over g}\)
      \(x\cos\alpha -\sin\alpha\)-4x^2{f\over g}\sin\alpha]
       + 2\(x\cos\alpha-\sin\alpha\)   \nonumber \\
&& A_{F_2 hh} = \sin2\beta [ \({A_g\over \mu} + 6{f\over g}\)
             \(x\sin\alpha +\cos\alpha\)+4x^2{f\over g}\cos\alpha]
            + 2\(x\sin\alpha+\cos\alpha\)  \nonumber\\
&& A_{\psi' hA} = \({A_g\over \mu} - 6{f\over g}\) {\(x^2+3\) \over x} 
\end{eqnarray}

If flatons produce a large number of flatinos, flatino decay into axions has 
to  has to be suppressed as well.  
Primary importance is the production of the lightest 
flatino $\tilde{F_1}$ which cannot decay into other flatons (axions)
or flatinos.
The flatino decay into ordinary particles comes from the superpotential
$W = {g\over \mpl} \hat{H_1}\hat{H_2}\hat{P}\hat{Q}$.
We find that the flatino decay into a Higgs and a Higgsino 
(more precisely, the lightest neutralino $\chi_1$)
has the rate;
\begin{eqnarray}
 \Gamma\(\tilde{F_i} \to \chi_1 h^0\) =
 {M^3_{\tilde{F_i}} \over 8\pi F_{\rm PQ}^2} {\mu^2\over M^2_{\tilde{F_i}} }
 \(x^2+9\)^2 C_{\tilde{F}_i}^2
 \( 1- {\(M_{\chi_1}+M_{h^0}\)^2 \over M^2_{\tilde{F_i}} }\)^{1/2}
 \( \(1+{M_{\chi_1} \over M_{\tilde{F_i}}}\)^2-
           {M^2_{h^0} \over M^2_{\tilde{F_i}}} \) \nonumber
\end{eqnarray}
where 
$C_{\tilde{F}_1}= (\sin\tilde{\alpha}+x^{-1}\cos\tilde{\alpha}) N_{\chi_1}$ 
and 
$C_{\tilde{F}_2}= (\cos\tilde{\alpha}-x^{-1}\sin\tilde{\alpha}) N_{\chi_1}$.
Here $N_{\chi_1}$ denotes the fraction of lightest neutralino in Higgsinos.

Let us now consider the flaton decay into ordinary fermions or sfermions.
The mixing terms between flaton and Higgs fields
allow a direct tree level coupling (after full mass matrix diagonalization)
between the usual fermions and flatons.
%
%
Parameterizing such a mixing with the parameter $\theta_{FH}$  the effective 
flaton-fermion interaction is $ h_f \theta_{FH}$
so that the rate of decay is
\be
\Gamma \(F_i \rightarrow f+\bar{f}\)= 
 N_c \frac{h_f^2  \theta_{FH}^2 } {16 \pi }M_{F_i} \sqrt{\(1-4\frac{m_f^2}
{M_{F_i}^2}\)^3}
\ee
where $N_c$ is a color factor for the fermion $f$.
Since  $\theta_{FH}\simeq \(\frac{v_{EW}}{F_{PQ}}\)$, for
  light fermions ($2m_f<M_F$)
$ \Gamma \(F_i \rightarrow f+\bar{f}\)/\Gamma \(F_i \rightarrow a+a\)\sim
h_f^2 v_{EW}^2/M_{F_i}^2 \sim m_f^2/M_{F_i}^2$ which is less than one.
Therefore, the rate of the flaton decay into ordinary fermion 
cannot be made sufficiently larger than  that into axions.

For the coupling between sfermions and flatons, we have two contributions.
One is a direct coupling coming from the scalar potential
\bear
&& V_{F\tilde{f}\tilde{f}} = {\mu \over \vpq} 
{\sqrt{x^2+9}\over x}{ v_1 \over \sqrt{2}}
 \( h_d  \tan \beta \tilde{D_L}^* \tilde{D_R}^*
+  h_e  \tan \beta \tilde{E_L}^* \tilde{E_R}^*
+  h_u   \tilde{U_L}^* \tilde{U_R}^*\)\\
&& \nonumber
\(F_1\(x \cos \alpha-\sin \alpha\)+ 
F_2\(\cos \alpha+x \sin \alpha\)\)+h.c.
\eear
where $\tilde{D}^*$ denote down-type squark, {\it etc}.

The other  arises  from  an indirect 
coupling induced by  the mixing between Higgs and flaton fields as
for the fermion case.
Taking in consideration the cubic soft A-terms we find
\be
V_{eff}=h_d A_d\, \theta_{F_i\, H_1}\, F_i \tilde{D_L} \tilde{D_R}
+  h_e A_e \, \theta_{F_i\, H_1}\, F_i  \tilde{E_L} \tilde{E_R}
+  h_u A_u \,\theta_{F_i\, H_2}\, F_i \tilde{U_L} \tilde{U_R}
\ee
so that effectively we have couplings of the size 
\be
G_{F\, sfermion}\sim  h_f (\mu+A_f){v\sub{EW} \over \vpq} 
\ee

%
Diagonalizing the sfermion mass matrix
we can write $\tilde{f}_R \tilde{f}_L=a_{11}\tilde{f}_1\tilde{f}_1 +
a_{22}\tilde{f}_2\tilde{f}_2 +a_{12}\tilde{f}_1\tilde{f}_2$
(where $a_{ii}\propto h_f$  so for $h_f \rightarrow 0$ we have 
$a_{12}  \rightarrow 1$).
Considering the decay of the light flaton we get
\be
\Gamma \(F_1 \rightarrow \tilde{f}_{i}+\tilde{f}_{j}\) \simeq 
N_c 
\frac{G_{F\, \tilde{f}}^2 }
{64 \pi\; M_{F_1}}
a_{ij}^2
\sqrt{ 1-4 \frac{m_{\tilde{f}}^2 }{M_{F_1}^2}} 
\ee
%
%
As observed in Ref.~\cite{eungjin}, the flaton may decay efficiently to
two light stops as $h_t \sim 1$ and $a_{ij}\sim 1$ and thus
a large splitting between light and heavy stops helps increasing the
flaton decay rate to light stops. 
This kind of mass splitting occurs also in the Higgs sector and furthermore
the light Higgs ($h^o$)
 is usually substantially lighter than the heavy Higgs ($H^o$)
in  the minimal supersymmetric standard model.  This should be contrasted to
the case with the mass splitting for stops  which 
requires some adjustment in soft parameters.
In this  paper, therefore, we will concentrate on the flaton decay into 
Higgses as a dominant mode of flaton decays.

\section{ Parameter space  analysis}

\label{5}

Our task is now to find the parameter space for which $B_a$ gets small 
enough. 
As a reference, let us try to see if $B_a<0.1$ (corresponding to 
$\delta N_\nu <0.7$) can be obtained  imposing the stronger condition that
$B_I<0.1$ for each $I=F_{1,2},\psi'$.  
We first note that the decay rates calculated in the previous sections are 
functions of the 4 variables $x=v_P/v_Q$, $f/g$, $A_f/\mu$ and $A_g/\mu$ 
disregarding their overall dependence on $F_{\rm PQ}$.

Before starting our discussion, let us make some remarks.
We are dealing with two kind of PQ models with a natural intermediate scale
1) the DFSZ  and  2) the  Hadronic model (KSVZ).  

$i)$ They have a common flaton potential and thus  the
same flaton   and flatino spectrum. But they have different interactions 
between flatons and matter.

$ii)$ The symmetries and  parameter space  constraints impose that the following
decays are forbidden  
$\psi'\rightarrow a a$, $F_1\rightarrow \psi' a$,
$F_1\rightarrow \psi'  \psi'$,
$F_1\rightarrow \tilde{F}  \tilde{F}$.

Neglecting for the moment  the model dependent flaton-matter interactions,
we  have to  analyze the  
decays
$F_{1,2}  \rightarrow a\; a$,
$F_{2}  \rightarrow \psi' a$ and the orthogonal 
$ \psi'\rightarrow  F_{1,2} \;  a$
plus the
flatons-flatinos interactions 
that through the chain 
$Flaton\rightarrow Flatinos \rightarrow Flatino\;-\; axion$
can also generate a non negligible axionic density at nucleosynthesis time.

The  decays of the flatons into two axions 
 doesn't have any  phase space suppression.
If we choose $\alpha=\alpha_1$ (see below Eq.~\ref{mixing})
 then we can suppress the 
rate $\Gamma \(F_{1} \rightarrow a+a\) $  only
taking $x^2=18$ and $
 \cos \alpha_1|_{x^2=18}=0.426 $
(giving for example  $\Gamma \(F_{2} \rightarrow a+a\) =3.6 
 \frac{M_{F_2}^3}{ 32 \pi F_{\rm PQ}^2}$). On the other hand, if
we choose $\alpha=\alpha_2$ we can suppress only
$\Gamma \(F_{2} \rightarrow a+a\) $
in the region $x^2=18$ and 
$\cos\alpha_2|_{x^2=18}=-0.905$
 (giving for example  $\Gamma 
\(F_{2} \rightarrow a+a\) =3.6  \frac{M_{F_1}^3}{ 32 \pi F_{\rm PQ}^2}$ ).

The  decay rate into a single axion  has the 
phase space suppression constraint
$m_{F_2}^2 \geq M_{\psi'}^2$ that translated in our parameters reads
$-\frac{g}{f}\frac{A_f}{\mu}\leq 12$.

When the flaton's branching ratio to flatinos becomes  sizable 
we  have also to impose
\be
 B\(I\to \tilde{F_2}\) B\(\tilde{F_2}\to a\tilde{F_1}\) < 0.1 \,.
\ee
with
$B\(I\to \tilde{F_2}\)=2\; 
B\(F_2\to \tilde{F_2}\tilde{F_2}  \)+
B\(F_2\to \tilde{F_2}\tilde{F_1}  \)+2\; B\(\psi' 
\to \tilde{F_2}\tilde{F_2}  \)+B\(\psi' \to \tilde{F_2}
\tilde{F_1}  \)$
and also the important requirement  that $\tilde{F_1}$ has to be heavier than
the lightest neutralino ($\chi_1^0$) and the light Higgs ($h^o$)  
since it has the unique decay mode to a neutralino and a Higgs.
Assuming $m_{\tilde{F_1}}$ is much larger than the
light Higgs mass, we impose a strong condition 
$ m_{\tilde{F_1}} > \mu $
that translates into the  approximate bound
$\left|{f\over g}\right|>{1 \over 3 x}$.

\subsection {Parameter space of KSVZ models}

As  discussed  already,
the light flaton $F_1$ can decay only into two gluons, 
so the simple requirement 
$B^{-1}_{F_1}=\Gamma[F_1 \rightarrow g g ]/ \Gamma[F_1 \rightarrow aa]\gg1$
impose  a quite strong  constraint on the
parameter space.
This ratio is given by 
\be
\frac{\Gamma \(F_1 \rightarrow g+g\)}{\Gamma \(F_1 \rightarrow a+a\)}=  
\frac{\alpha^2_S\(M_{F_1}\)}{ \pi^2}
\frac{4}{9} N_E^2 \(\frac{ \sin \alpha \(x^2+9\)}
{x\(9\cos \alpha-x \sin \alpha\)}  \)^2
\(1+\frac{95}{4}\frac{\alpha_S\(M_{F_1}\)}{ \pi}\) \,.
\ee
In order to have a large  number  
we can use only  the $x$ parameter 
and the  number of exotic quark ( to be  as large as possible).
For $\alpha_S\(M_{F}\)\simeq 0.1$, 
this ratio  can be  larger than 10  accepting a tuning of the $x$ parameter
as follows 
\bea
\sqrt{18} -0.04  < & x &< \sqrt{18}+0.04 \quad \mbox{for}\quad N_E=1, 
        \nonumber\\
\sqrt{18} -0.3  < & x &< \sqrt{18}+0.3    \quad\mbox{for} \quad N_E=9.
\eea
However, in such region we don't  find any solution in which, 
 $B_a(F_2)$ and  $B_a(\psi')$ are less than 0.1  at same time.
Therefore the KSVZ model cannot give a satisfactory solution of the
nucleosynthesis bound unless
some extra fine tuning on the 
initial densities of the flatons $F_2,\,\psi'$ 
can be dynamically achieved.

\subsection {Parameter space of DFSZ models }

In the DFSZ model the number of the  possible decay
channels is much larger than the other model 
and it is not an easy task to find, in the four parameter space, some 
easy understandable avaliable region.
To be as independent as possible of the 
soft supersymmetry breaking parameters 
we will try to make analytic computations on the 
rates of the flaton decays into
Higgs particles, in particular into the lightest 
Higgs ($h^0$) whose mass has an
automatic upper bound of $\sim 140$ GeV \cite{quiros}. 
We will concentrate on the region with
 $\frac{f}{g}$ negative and $ \left|\frac{f}{g} \right|\ll 1$ and $x\gg1$
which turns out to be required for $B_a<0.1$.

To open the decay channels of the flatons into Higgs particles we have,
 in particular, to  impose 
 $M_{\psi'} > M_A >0$ which
requires (from now on we will use the unequal symbols as strictly satisfied)
 $\;\;\;\frac{A_g}{\mu}< \left|\frac{f}{g} \right|\,x^2 
\;\;\;$  with
\be
\left|\frac{f}{g}\frac{A_f}{\mu} \right|\,x^2>2\,\left
|\frac{A_g}{\mu}+\frac{f}{g}\,x^2 \right| \,.
\ee
For simplicity we divide into two regimes
\be
|\frac{A_g}{\mu}|< |\frac{f}{g}|\,x^2 \,,
 \;\;\;\;\;\;\;\left|\frac{ A_f }{ \mu} \right|> 2 \label{reg1}
\ee
and
\be
 \left|\frac{A_g}{\mu}\right|> \left|\frac{f}{g}\right|\,x^2 \,,
 \;\;\;\;\;\;\;
\left|\frac{ A_g }{ A_f} \right|< {1\over2} \,x^2 \left| \frac{f}{ g}\right|\,,
\;\;\;\;\;\;\;
\left| \frac{A_f }{ \mu} \right|> 2 \label{reg2} \,.
\ee
Remember that the positivity of flaton masses requires
\be
 \left|\frac{ A_f }{ \mu} \right|<  x^2 \left|\frac{ f}{g}\right| \,.
\ee
Combining altogether we get 
$  \left|\frac{f}{g}\right|>\frac{1}{2\,x^2}$ in the regime (\ref{reg1})  and
$  \left|\frac{f}{g}\right|>\frac{2}{\,x^2}$ in the regime (\ref{reg2}).
Besides, if flatino production rates are sizable, we also have to impose
$R_{\tilde{F}_2} \equiv \Gamma(\tilde{F}_2 \to \chi_1 h^0)/
 \Gamma(\tilde{F}_2 \to a \tilde{F}_1)
 > 10$ and 
$M_{\tilde{F}_1}>M_{\chi_1}+M_{h^0}$ to open the decay mode
$\tilde{F}_1 \to \chi_1 h^0$.  The condition $R_{\tilde{F}_2}>10$ gives
the restriction $
 \left| f/g\right|<0.02 \,x\, N_{\chi_1}   $  and $ |\frac{f}{g}|>
\frac{1}{3\,x}$.

Then we study, in our limit, the constraints given by the conditions 
$R_{\psi'}\equiv\Gamma(\psi' \to h^0 A^0)/\Gamma(\psi'\to aF_1) > 10$ and 
$R_{F_i}\equiv\Gamma( F_i \to \to h^0 h^0)/\Gamma(F_i\to a a ) > 10$. 
The ratios $R$ are
\bea
R_{\psi'}&\sim&
 \frac{1}{144}\left(\frac{g}{f}\right)^2
\frac{\mu^2}{A_f^2}\left(\frac{A_g}{\mu}-
6\, \frac{f}{g}\right)^2 \label{condi1} \nonumber
\\
R_{F_1} &\sim&  \frac{1}{ 4 }\frac{\mu^4}{M_{F_1}^4}
(\frac{A_g}{\mu}-2 \,\frac{f}{g}\,x^2+2)^2\label{condi2}
\\ 
R_{F_2} &\sim&  10^{-3}\, x^4 \frac{\mu^4}{M_{F_2}^4} \;(\frac{A_g}{\mu}
+18 \,\frac{f}{g}\,
+2)^2\label{condi3}  \nonumber
\eea
where  
$M_{F_1}^2\sim |\frac{f}{g} \frac{A_f}{\mu}| x^2 \mu^2$  and
$M_{F_2}^2\sim 12\,\frac{f^2}{g^2}  x^2 \mu^2$
for $ \frac{A_f}{\mu}< 12 |\frac{f}{g}|$, 
and $M_{F_1}\leftrightarrow M_{F_2} $ for $ \frac{A_f}{\mu}>
 12 |\frac{f}{g}|$. 
We can now divide the parameter space into four regions. 

I) $\left| \frac{A_g}{\mu}\right|>\left| \frac{f}{g}\right|\, x^2
   \;\;\;\;\;\;$,
$I\!I$) $2<\left| \frac{A_g}{\mu}\right|<\left| \frac{f}{g}\right| \,x^2
         \;\;\;\;\;\;$,
$I\!I\!I$) $ \left| \frac{f}{g}\right| <\left| \frac{A_g}{\mu}\right|<2
        \;\;\;\;\;\;$,
$I\!V$) $  \left| \frac{A_g}{\mu}\right|<\left| \frac{f}{g}\right|
           \;\;\;\;\;\;$.
Depending on $ \frac{A_f}{\mu}< 12 |\frac{f}{g}|$ or $ \frac{A_f}{\mu}>
12 |\frac{f}{g}|$  we will have region ${\it a}$ or region $ {\it b}$.

We find that all of the ${\it a} $ 
regions are forbidden, and so is the $I\!V_b$ region.
The constraints for the regions are as follows.
\bea
&I_b& x>14,\; \; \;  A_g<0,\; \; \; 
12 \left|\frac f g\right| < \left|\frac{A_f} \mu \right|
 \nonumber\\
&& \;\; 1< \frac12
\left|\frac {A_f}\mu\right|< \left|\frac f g\right|\frac{x^2} 2
<\left|\frac{A_g}\mu\right| < 2 \(  \left|\frac f g \right| 
\frac {x^2} 2 \)^2 \nonumber\,.\\
&I\!I_b& x>9,\; \; \; A_g>0,\; \; \;
\left| \frac{f}{g}\right|<3\times 10^{-2} \nonumber\\
&&\;\;
2<\left| \frac{A_f}{\mu}\right|<
 \left| \frac{f}{g}\right| \,x^2,
\;\;\;
2<\left| \frac{A_g}{\mu}\right|<
\left| \frac{f}{g}\right| \,x^2 \,.\nonumber\\
& I\!I\!I_b& A_g>0,\; \; \;
1 < 2\left|\frac f g\right| x^2,\nonumber\\
&&\;\;\;
2\left|\frac f g\right| < \left|\frac{A_f}\mu\right| \left|\frac f g\right|
 < 2\times 10^{-2},
\;\;\;\;
\left|\frac f g\right| < \left|\frac {A_g}\mu\right| < 2 \,.
\eea

To summarize, we find that:\\
- In all cases, $x$ has to be large ($\gsim 10$)\\
- In cases $I\!I$ and $I\!I\!I$,
$\left| \frac{f}{g}\right| $ has to be small ($<3\times 10^{-2}$)
but it has no useful upper bound in case $I$. In all cases
$x^2\left|\frac f g\right|\gsim 1$.\\
- In all cases, $|A_f|\sim |A_g|\sim|\mu|$ is a possibility.

\section{Conclusions}
\label{6}

We have explored the cosmology of a particularly attractive extension of 
the Standard Model, which has a Peccei-Quinn symmetry broken only by
two `flaton' fields $\phi_P$ and $\phi_Q$, 
characterized by a very large vev ($10^{10-12}$ GeV)
 and a relatively small mass
 ($10^{2-3}$ GeV).
 These and their superpartners generalize the
saxion and axino, that in the non-flat case are the only fields with 
soft masses.

In contrast with more general models the density of 
dark matter axions can be estimated with essentially no assumption about
other sectors of the theory, assuming only that 
$\phi_P$ has a positive effective  mass-squared in the early Universe.
If $\phi_Q$ also has a positive effective mass-squared the axion
is an excellent dark matter candidate. In the opposite case
the axion density is probably to low in this particular model.

Our main concern has been with a different, 
highly relativistic, axion population that is produced by 
flaton decay. We have calculated the rates for all relevant channels
and  examined the constraint 
that the energy density of these axions does not upset the
predictions of the standard nucleosynthesis.

For the KSVZ  model we find that
it is almost impossible to satisfy such a  cosmological  bounds
due to the fact that the only  flatonic decay channel in competition with 
the axionic one is into gluons
(through axial anomaly) which is too much constrained. 

 For the DFSZ  model  there are more decay channels.
To evade complicated phase space suppressions 
we concentrate on the Higgs decay products, as  
the mass of the lightest Higgs has naturally
a relatively low upper bound and the mass of the other Higgs and flaton 
fields are fixed by the parameters of the flatonic potential itself.
We have quantified a portion of the parameter space available  
showing the strength of this  model.

An interesting question, lying beyond the present investigation,
is whether the allowed region of parameter space
can be achieved in a supergravity model with 
universal soft parameters.

\newcommand\pl[3]{Phys. Lett. {\bf #1}, #2 (19#3)}
\newcommand\np[3]{Nucl. Phys. {\bf #1}, #2 (19#3)}
\newcommand\pr[3]{Phys. Rep. {\bf #1}, #2 (19#3)}
\newcommand\prl[3]{Phys. Rev. Lett. {\bf #1}, #2 (19#3)}
\newcommand\prd[3]{Phys. Rev. D{\bf #1}, #2 (19#3)}
\newcommand\ptp[3]{Prog. Theor. Phys. {\bf #1}, #2 (19#3)}

\end{document}